\documentclass[intlimits,twoside,a4paper]{article}

\usepackage[cp1251]{inputenc}
\usepackage{multirow}
\usepackage[eqsecnum]{cmpj3}

\usepackage{bm}
\newcommand{\vw}{{\bf v}}
\newcommand{\vq}{{\bf q}}
\newcommand{\vj}{{\bf j}}

\newcommand{\vR}{{\bf R}}

\issue{2023}{26}{3}{33601}
\doinumber{10.5488/CMP.26.33601}

\title[An ab initio study of some liquid 5d transition metals]%
{An ab initio study of the static, dynamic and electronic 
properties of some liquid 5d transition metals near melting}

\author{D. J. Gonz\'alez\orcid{0000-0003-1116-3473},   L. E. Gonz\'alez\orcid{0000-0001-6264-8329} } 
\address{ 
Departamento de F\'\i sica Te\'orica,  
Universidad de Valladolid, 47011 Valladolid, Spain}


\date{Received January 16, 2023, in final form February 17, 2023}

\begin{document}

\maketitle

\begin{abstract}
	We report a study on the static and dynamic properties of several 
	liquid 5$d$ transition  metals at thermodynamic conditions 
	near their respective melting points. 
	This is performed by resorting to 
	{\em ab initio} molecular dynamics simulations in the framework of 
	the density functional theory. 
	Results are presented for the 
	static structure factors and pair distribution functions; moreover, 
	the local short range order in the liquid metal is also analized. 
	As for the dynamical properties, 
	both single-particle and collective properties
	are evaluated. 
	The dynamical structure shows the propagating density fluctuations, 
	and the respective dispersion relation is obtained. 
	Results are also obtained for the 
	longitudinal and 
transverse current spectral functions along with the associated dispersion of 
collective excitations. For some metals, we  found the existence of two branches of 
transverse collective excitations in the region around the main peak of the
	structure factor. 
Finally,  several transport coefficients are also  calculated.

\printkeywords
\end{abstract}

\section{Introduction}

This paper reports an {\it ab initio} molecular dynamics (AIMD) 
simulation study on the 
static, dynamic and electronic properties of several 
liquid 5$d$ transition metals (Hf, Ta, W, Re, Os, Ir, Pt and Au) 
at thermodynamic conditions just above melting at room pressure. 

Except Au, all the other metals   
belong to the group of the  
so-called refractory metals which are characterized by 
very high melting points ($\geqslant$ 2000 K). 
They also share some properties such as 
high density, retention of mechanical strength at 
high temperatures,  high hardness at room temperature and 
resistance to the damages of corrosion, wear and deformation.  
Because of their exceptional high-temperature properties, 
the refractory metals and alloys are a good choice   
 for high-temperature applications, e. g., those materials used for  
fission and fusion reactors (first wall, blanket and divertor).  

Other interesting applications of these metals have to do with 
defense and aerospace industries, medical devices, low-temperature 
superconductor applications, manufacture of very hard-wearing and 
durable electrical components, chemical industry, etc.
Consequently, all those above mentioned  technological applications 
justify the research effort aimed at achieving  
a better understanding of their structural, dynamical 
and electronic properties. 

In this paper we are focused on the molten state, which 
because of  the high melting temperatures  and  
high reactivity  of these refractory metals, poses important challenges when 
trying to measure their liquid state properties. 
Nevertheless, a good understanding of their thermophysical properties is 
important for the studies on phase transformations, nucleation, atomic dynamics and 
surface physics as well as industrial processes (i.e., refining, casting and welding)  
and for designing alloys.

Notwithstanding the technological relevance of these metals, there is a 
scarcity of experimental data concerning their static, dynamic 
and electronic properties in the molten state; obviously, this is related 
to the challenges posed by their high melting temperatures.

Two basic magnitudes related to the structural short-range order in a liquid metal are 
 the static structure factor, $S(q)$, and the 
pair distribution function, $g(r)$.  
In the case of the liquid 5$d$ metals considered in this work, it is no wonder that the available 
experimental structural data are for the two metals with the lowest melting temperature, namely, 
Pt and Au. In fact, these data were obtained, more than forty years ago, by 
Waseda and coworkers \cite{WasBook} that performed X-ray 
diffraction (XD) measurements of the $S(q)$ of liquid
l-Pt and l-Au at 
thermodynamic conditions near their respective melting points. Since then, no additional 
measurements  of the $S(q)$  
have been performed for any of these liquid metals with the  exception of  l-Ta for which 
XD experiments have been recently performed under thermodynamic conditions of 
internal negative pressure \cite{Katagiri} and results have been obtained 
for its $S(q)$  in a pressure range from  5.6 GPa to  2.7 GPa.

As for the dynamical properties, we mention the 
experimental study by Guarini  et al. \cite{AuGuarini} on 
the microscopic dynamics of l-Au at $T=1373$ K. Specifically, they 
performed inelastic neutron scattering (INS) measurements, and the  
dynamic structure factor, $S(q, \omega)$, was determined 
within the range $0.6 \leqslant q \leqslant 1.6$ \AA$^{-1}$. 
The propagating collective excitations were detected and its associated dispersion 
relation yielded an adiabatic velocity of sound in good agreement
with the experimental value of $\approx 2568$ m/s \cite{Blairs07}. 
We are not aware of any similar study performed for any of the other 5$d$ metals 
considered in this work. 
In the same study, the authors also performed AIMD simulations that showed 
good agreement with the experimental data.

The additional available experimental data refer to the 
transport coefficients only. 
The sound velocity was measured, in addition to l-Au, for l-Ta, l-W and 
l-Pt \cite{Blairs07}.
The shear viscosities were measured by 
Ishikawa et al. \cite {IG15,Ishi03,Ishi12,Ishi13} by means of  
levitation techniques;  however, 
their diffusion coefficients have not been determined yet.

This shortage of experimental data 
for this group of metals highlights the 
importance of resorting to  other approaches such as 
theoretical and computer simulation methods to 
extract information about their static, dynamic and electronic 
properties.

However, few theoretical studies, 
either based on semiempirical or more fundamental methods, 
have been devoted to the study of these   
liquid 5$d$ transition metals  so far  
and they have mainly focused on thermodynamic and static structural 
properties.
There are a few studies based on the effective potentials for l-Pt and l-Au, by
Alemany  et al. \cite{Alemany,Alemany2} and by Gosh  et al. \cite{Gosh},
where the static structure and some transport properties were evaluated.
Using a more fundamental approach
Bhuiyan  et al. \cite{Bhuiyan12} performed a MD study of 
several static and dynamic properties of l-Au near melting 
by resorting to the orbital-free ab initio MD simulation 
method (OF-AIMD), which is based on the Hohenberg and Kohn version \cite{HK} of the 
Density Functional Theory (OF-DFT), where 
the basic variable is the total valence electronic density. They obtained good results for the 
static structure as well as for some transport coefficients such as the diffusion coefficient and 
the shear viscosity. 
In addition to those already mentioned for l-Au \cite{AuGuarini}, we finally mention AIMD simulations 
based on the Kohn-Sham formulation of DFT (KS-DFT) \cite{KS}
that were performed for l-Ta and undercooled l-Ta by Jakse  et al. \cite{Jakse,Jakse2}, where the 
static structure was studied, and those for l-Pt by 
del Rio  et al. \cite{RioGG} that 
evaluated
several transport properties and static and dynamic structural properties, reporting good
agreement with experiments. 
Some of those results for l-Pt are included in this work for completion.

The AIMD simulation study presented in this paper, is based on 
the KS-DFT formalism.   The liquid 
metal is modelled as an interacting system of ions and electrons  
and for any  ionic configuration, the associated electronic ground state 
is evaluated by means of the KS-DFT approach. Then, the  
forces acting on the ions are obtained via the Hellmann-Feynman theorem,  
and their positions evolve according to
classical mechanics while the electronic subsystem 
follows adiabatically. The present AIMD simulation method requires  
large computational capabilities and imposes  severe limitations concerning  
the simulation times and size of the systems under study;  
however, these disadvantages  are somewhat balanced by the 
accuracy of the obtained results.

The paper is structured as follows: the next section  
summarizes the basic ideas underlying  the AIMD simulation 
method along with some technical details.
In section \ref{results1} we report the results of the calculations 
which are compared with the available experimental data along with 
some discussion. Finally, a brief summary and 
conclusions are given in section \ref{conclusions}. 
\newpage

\section{Computational method}
\label{Cmethod}

Table \ref{states} lists the specific $5d$ metals and the corresponding 
thermodynamic states for which the AIMD simulation study is performed. 
The simulations were carried out by using a cubic cell 
containing 120 atoms (150 atoms in l-Au).

The AIMD simulations were implemented using  the DFT based  
Quantum-ESPRESSO package~\cite{espresso,espresso2}. 
Within this framework, the electronic exchange-correlation 
energy was described by the 
generalized gradient approximation of Perdew-Burke-Ernzerhof  
\cite{PBE}, except l-Pt and l-Au for which we used the local density approximation 
of Perdew and Zunger \cite{PZ}. On the other hand, 
the ion-electron interaction was accounted for by means of 
an ultrasoft pseudopotential \cite{vanderbilt}, which was  
generated from a scalar-relativistic calculation and it included  
non-linear core corrections. 
The only exception was l-Ir for which
we  used a norm-conserving  Troulliers-Martins type pseudopotential.
Table \ref{states} specifies, for each metal, the number of valence electrons 
which are explicitly considered in the calculation. 
Note that in the cases of Hf through Os the $5s$ and $5p$ electrons
are included as semi-core states.

The simulations started with the ions placed at some random 
initial positions within 
the cell and then the system was thermalized 
during $5-10$ ps of simulation time; 
therefrom, microcanonical AIMD simulations, with a time step 
of 0.0055 ps (0.0075 ps for l-Au), were
performed over the number of time steps given in table \ref{states}. 
We used a plane-wave representation with an energy cutoff within the range from 
25 to 40 Ry for the wavefunctions and 150 to 350 Ry for the electronic density.
The single $\Gamma$ point was used for sampling 
the Brillouin zone. 

The number of equilibrium configurations listed in 
table \ref{states} were those used  
for the evaluation of the static, dynamic and electronic
properties of the corresponding liquid metal. 
We notice that this same simulation method has already proved 
its capability to deliver an accurate description of several static, dynamic and transport 
properties of other bulk liquid metals 
\cite{CGGLHgCd,CGGLHgCd2,CGGLHgCd3,CGGLHgCd4,RRGGTiNiFe,RGGNi,RGGNi2,Bea_3d,Bea_3d2,Gonzalez_4d}.

\begin{table}[!htb]

\caption{
Thermodynamic input data of the liquid $5d$ transition metals considered in 
the present AIMD simulation study, where  $\rho$ is the total ionic number
density (taken from \cite{IG15}), $T$ is the temperature, 
$N_{\rm part}$ is the number of particles in the simulation box, 
$Z_{\rm val}$ is the number of 
valence electrons in the pseudopotential and 
$N_c$ is the total number of generated configurations. \label{states}} 
\vspace{2ex}
\begin{center}
\begin{tabular}{cccccccc}
& $\;\;\;$  &  $\rho$ (\AA$^{-3}$)  & $T$(K) & $N_{\rm part}$ 
	& $Z_{\rm val}$ &  $N_c$  \\
\hline
	& Hf & 0.0382 &  2550  &120 & 12 & 12000  \\
	& Ta & 0.0473  &  3350  &120&  13 & 13000   \\
	& W & 0.0562  &  3750  & 120 & 14 & 13000  \\
	& Re & 0.0591  &  3600  & 120 & 15 & 10500  \\
	& Os & 0.0605  & 3400  & 120 &  16 & 11500  \\
	& Ir & 0.0623  & 2750  & 120 & 9 & 20000  \\
	& Pt & 0.0577  & 2053  & 120 & 10 & 19500  \\
	& Au & 0.0526  & 1423  & 150 & 11 & 20000  \\
\hline
\end{tabular}
\end{center}
\end{table}


\section{Results and discussion}
\label{results1}

\subsection{Static properties}
\label{results}

\begin{figure}[!htb]
\centerline{\includegraphics[width=85mm,clip]{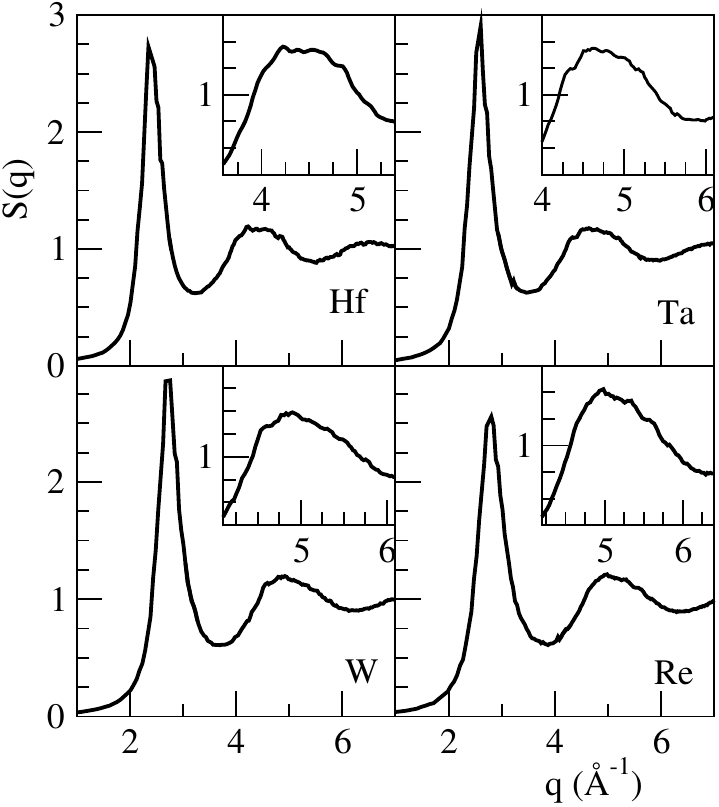}}
\caption{Static structure factor, $S(q)$, of l-Hf, l-Ta, l-W and l-Re. 
Continuous line: present AIMD calculations.  
The inset shows a closer view of the second maximum. }
\label{fig_sqA}
\end{figure}
\begin{figure}[!htb]
\centerline{\includegraphics[width=85mm,clip]{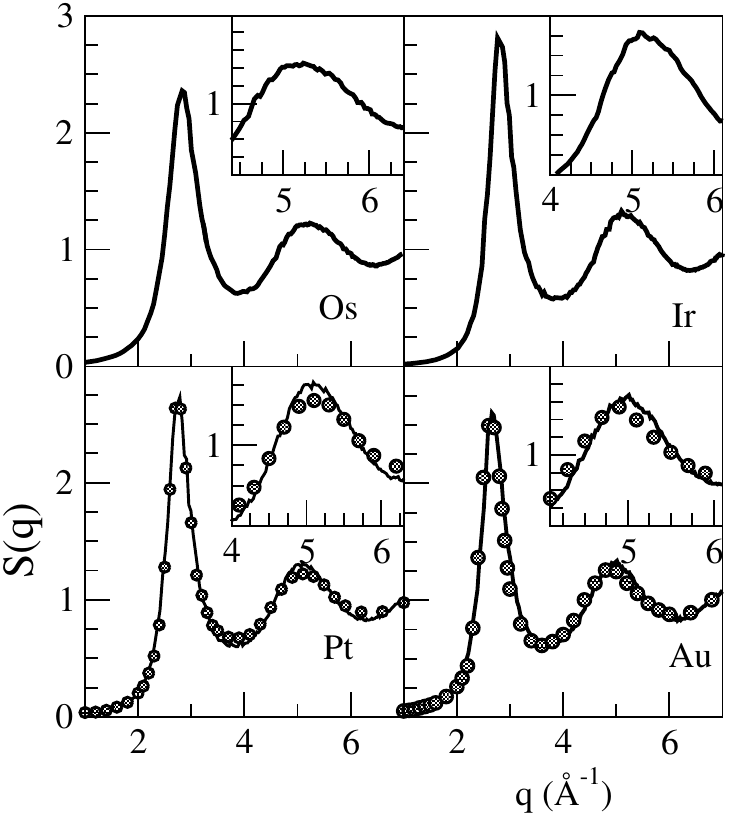}}
	\caption{Same as the previous graph but for l-Os, l-Ir, l-Pt \cite{RioGG} and l-Au. 
Symbols: XD data from Waseda's group \protect\cite{WasBook,WasOht}. 
	}
\label{fig_sqB}
\end{figure}

The results obtained for the respective  static structure factors, $S(q)$, are  
plotted in figures~\ref{fig_sqA}--\ref{fig_sqB}, where they are compared with the 
(scarce) available experimental data. We observe that for both l-Pt and l-Au there is a 
very good agreement, both in the position and amplitude of the 
oscillations, with the corresponding XD data. 
As for the other $5d$ liquid metals considered in this study,  there are no structural 
data available for comparison. In the case of l-Ta, the 
recent XD measurements refer to thermodynamic states different from the present AIMD 
simulations.

Figures \ref{fig_sqA}--\ref{fig_sqB} also provide a closer look at the 
shape of the second peak of the $S(q)$. We obtain a noticeable asymmetric shape  
in the group from l-Hf to l-Re  but it  
 becomes more symmetric for l-Os, l-Ir and l-Au. 
We mention that an asymmetric shape of the second peak with a shoulder  
 on its high-$q$ side  was experimentally found  in several liquid metals, 
including some transition metals (Ti, Fe, Ni) \cite{SchenkPRL,LeePRL04,Kelton-TiZrNi} 
and it was interpreted as an indicator of a noticeable existence of
icosahedral local order in the liquid.

We  used the low-$q$ values of the calculated $S(q)$, i.e., the values within 
the range $q$ $\leqslant$ 1.2 \AA$^{-1}$, 
to obtain a rough estimate  for $S(q \to 0)$. This is achieved by  
using a least squares fitting of $S(q)=s_0+s_2q^2$, and the results are 
given in table \ref{static}. Then, the relation   
where $k_{\text{B}}$ is Boltzmann's constant, allows to determine the associated 
isothermal compressibility, $\kappa_T$, and the results are given in  
table \ref{static}. 
We  additionally included the values suggested by 
Marcus \cite{Marcus17}, which were obtained by using semiempirical 
expressions which involve the measured and/or estimated values of  
several other thermophysical magnitudes, and also the experimental
values reported by Blairs \cite{Blairs07} for l-Ta, l-W, l-Pt and l-Au.

\begin{table}[!htb]

\caption{
The calculated values for $r_{\rm min}$ (in \AA),  
coordination numbers CN, $S(q \to 0) $ and isothermal 
compressibilities  $\kappa_T$ (in 10$^{-11}$ Pa$^{-1}$ units) 
for the liquid $5d$ transition metals at 
the thermodynamic states given in table \ref{states}. 
	The results for l-Pt are taken from \cite{RioGG}.
	The numbers in parenthesis 
	are semiempirical estimates from Marcus \cite{Marcus17}. 
	Experimental data for $\kappa_T$ at the melting temperature
	are taken from \cite{Blairs07}. \label{static} }
\vspace{2ex}
\begin{center}
\begin{tabular}{cccccc}
\hline
 $\;\;\;$  & $r_{\rm min}\;\;\;$ & CN $\;\;\;$ & $S(q \to 0)\;\;\;\;$ & 
	$\kappa_T\;\;$ & $\kappa_T^{\rm exp}$ \\
\hline
	Hf & 4.23 & 12.2 &   0.024 $\pm$ 0.002 & 1.78 $\pm$ 0.20  (1.78) & \\
	Ta & 3.95 & 12.2 & 0.022 $\pm$ 0.002 & 1.01 $\pm$ 0.15 (1.68) & 1.00 \\
	W & 3.78 & 12.9 & 0.016 $\pm$ 0.002 & 0.55 $\pm$ 0.10   (0.99) & 0.95 \\ 
	Re & 3.62 & 12.1 & 0.015 $\pm$ 0.001 & 0.51 $\pm$ 0.05   (0.92) & \\
	Os & 3.51 & 11.3 & 0.014 $\pm $ 0.001 & 0.49 $\pm$ 0.05  (0.86) & \\ 
	Ir & 3.50 & 11.8 & 0.009 $\pm$ 0.001 & 0.36 $\pm$ 0.05 (0.82 ) & \\
	Pt & 3.77 & 12.8 & 0.011 $\pm$ 0.001 & 0.67 $\pm$ 0.05 (1.02 ) & 0.84  \\
	Au & 3.63 & 11.2 & 0.012 $\pm$ 0.001 & 1.09 $\pm$ 0.05 (1.19 ) & 1.31 \\
\hline
\end{tabular}
\end{center}
\end{table}

\begin{figure}[!htb]
\centerline{\includegraphics[width=85mm,clip]{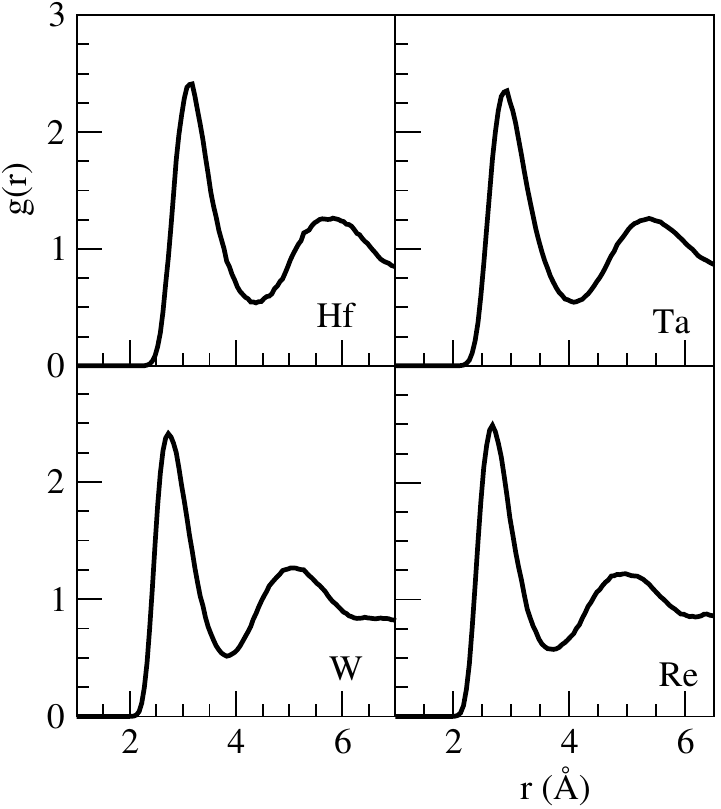}}
\caption{Pair distribution function, $g(r)$, of l-Hf, l-Ta, l-W and l-Re. 
Continuous line: present AIMD calculations.  
	} 
\label{fig_grA}
\end{figure}

\begin{figure}[!htb]
\centerline{\includegraphics[width=85mm,clip]{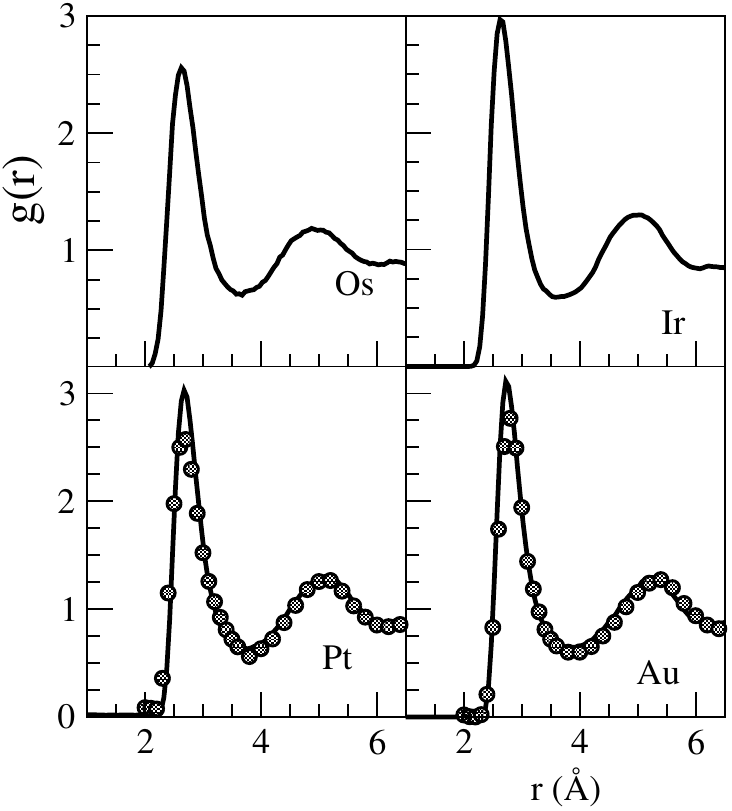}}
\caption{Same as the previous graph but for l-Os, l-Ir, l-Pt and l-Au. 
Symbols: XD data from Waseda's group \protect\cite{WasBook,WasOht}. 
	} 
\label{fig_grB}
\end{figure}

The pair distribution function, $g(r)$, provides some insight into the 
short range order in the liquid metal and  
figures \ref{fig_grA}-\ref{fig_grB}  depict the obtained AIMD results  
along with the available experimental data. 
As it happened with the $S(q)$, we observe a very good 
agreement for both l-Pt and l-Au.  
The average number of nearest neighbors (also called coordination number, CN) 
around any given ion can be evaluated by 
integrating the radial distribution function,    
$ 4 \piup \rho r^2 g(r)$, up to the position of its first 
minimum, $r_{\rm min}$. 
Table \ref{static} lists the obtained results which are 
similar to those calculated for other simple liquid 
metals near melting  \cite{Balubook}.

A more thorough description of the short range order in  a liquid metal  
is provided by the common neighbor analysis \cite{Andersen,Andersen2,Andersen3} (CNA) method. 
It gives three-dimensional information 
	about the ions surrounding each pair of 
	 ions which are near neighbors.  
	Each pair is characterized by four 
	indices which allow to discern among different local  
	 structures like fcc, hcp, bcc, and icosahedral environments. 
Thus, the fcc order only has 1421-type pairs,  
the hcp structure  has the same number of 1421 and 1422-type pairs,  
the bcc order has six 1441-types pairs and eight 1661 type-pairs and the  
13-atom icosahedron (ico) has twelve 1551 type-pairs. The   
 deformation of the regular 1551-type structure (i.e., when a bond 
is broken)  yields the 
1541 and 1431-type pairs and therefore their presence points to 
the existence of a somewhat distorted icosahedral order. 
Finally the 1311 and 1321-type pairs are related to disorderly structures.
For more details about the CNA method we refer 
to references \cite{Andersen,Andersen2,Andersen3,Bea_3d,Gonzalez_4d}.  
\begin{table}[!htb]

\caption{Common neighbour analysis of the AIMD configurations
obtained for the liquid $5d$ transition metals at 
the thermodynamic states given in table \ref{states}. For comparison 
	we include the values for some 
	common local structures.
	The data for l-Pt are taken from \cite{RioGG}. \label{cna}}
\vspace{2ex}
\begin{center}
\begin{tabular}{cccccccccccccc}
\hline
	Pairs & & 1551 & 1541 & 1431 &  1421 & 1422 & 1311 & 1321 &  1441 & 1661  \\
\hline
	l-Hf &  & 0.28 & 0.21 & 0.22 & 0.04 & 0.05 & 0.02 & 0.03 &  0.08 & 0.10  \\
	l-Ta &  & 0.32 & 0.19 & 0.21 & 0.02 & 0.05 & 0.02 & 0.03 &  0.06 & 0.09  \\
	l-W &  & 0.30 & 0.19 & 0.19 & 0.02 & 0.03 & 0.02 & 0.03 &  0.06 & 0.09  \\
	l-Re &  & 0.18 & 0.19 & 0.24 & 0.06 & 0.10 & 0.06 & 0.04 &  0.03 & 0.04  \\
	l-Os &  & 0.17 & 0.18 & 0.23 & 0.05 & 0.09 & 0.09 & 0.07 &  0.03 & 0.04  \\
	l-Ir &  & 0.17 & 0.22 & 0.23 & 0.06 & 0.09 & 0.04 & 0.04 &  0.04 & 0.05 \\
	l-Pt &  & 0.14 & 0.21 & 0.22 & 0.07 & 0.10 & 0.06 & 0.05 &  0.05 & 0.04 \\
	l-Au &  & 0.10 & 0.17 & 0.23 & 0.07 & 0.14 & 0.12 & 0.06 &  0.03 & 0.03 \\
\hline
	HCP & & 0.0 & 0.0 & 0.0 & 0.50 & 0.50 &  0.0 & 0.0 & 0.0 & 0.0  \\
	FCC & & 0.0 & 0.0 & 0.0 & 1.0 & 0.0 & 0.0 & 0.0 &  0.0 & 0.0  \\
	BCC & & 0.0 & 0.0 & 0.0 &  0.0 & 0.0 & 0.0 & 0.0 & 0.43 & 0.57  \\
\hline
\end{tabular}
\end{center}
\end{table}
\begin{figure}[!htb]
\centerline{\includegraphics[width=85mm,clip]{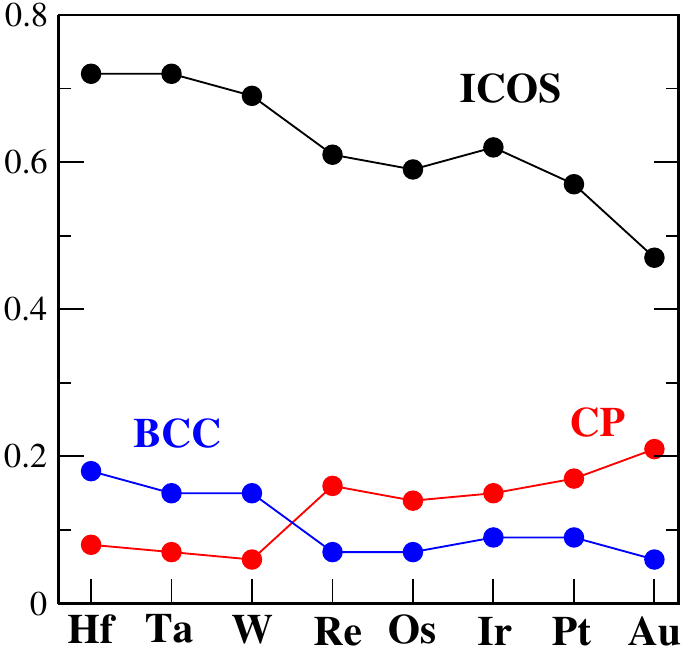}}
\caption{(Colour online) Variation of the 
most abundant bonded pairs.}
\label{fig_Cna}
\end{figure}
Table \ref{cna} and figure \ref{fig_Cna}  
summarize the results obtained in this work. 
The five-fold symmetry dominates in all these metals 
because the sum of perfect and distorted  
ico-structures ranges between  $\approx$ 50\% (l-Au) and 
$\approx$ 72\% (l-Hf, l-Ta) of the pairs. The amount of  
 local bcc-type pairs is significant in l-Hf, l-Ta and l-W whereas it is 
small for the other metals which, on the other hand, show an important  
	amount of fcc and hcp-type pairs (from 
	$\approx$ 12\% in l-Pt  to $\approx$ 21\% in l-Au ). Moreover, this 
	group of metals also shows a noticeable fraction of 
	disordered 13xx(1311 + 1321)-type pairs  (from 
	$\approx$ 8\% in l-Ir  to $\approx$ 18\% in l-Au). 

Our results for l-Ta are statistically equivalent to the early ones obtained
by Jakse  et al. in 2004 \cite{Jakse,Jakse2} above the melting temperature. 
We do not discern hints of the A15 structure that they observed in the
undercooled liquid, because the ratio 1661 to 1441 is very similar to the one
corresponding to bcc structures. Nevertheless, this signature was very dependent
on the density of the undercooled liquid, so it is not surprising that 
it is absent in our liquid sample.
It is also worth mentioning that the secondary structures present in the 
liquid $5d$ metals correlate well with the phases from which the corresponding 
solids melt, namely, bcc in the cases of Hf, Ta and W, and close-packed in 
the rest of cases (hcp for Re and Os, fcc for Ir, Pt and Au).

\subsection{Dynamic properties}

We have evaluated several dynamical properties obtained from time dependent
correlation functions. Some of these functions depend on 
the wave-vector $\vq$,
but the isotropic behavior of the fluid allows to reduce such
dependence to a dependence on $q \equiv| \vq |$ only.

\subsubsection{Single particle dynamics}

\noindent

The (normalized) velocity autocorrelation function (VACF)
of a tagged ion in the fluid, $Z(t)$, is defined as 

\begin{equation}
Z (t) = \langle \vw_i(t) \cdot \vw_i(0) \rangle
/ \langle v_i^2 \rangle , 
\end{equation}

\noindent with $\vw_i(t)$ being  the velocity of ion $i$ at time $t$, and the 
angular brackets denote an average over particles and time origins. 

\begin{figure}[!htb]
\centerline{\includegraphics[width=85mm,clip]{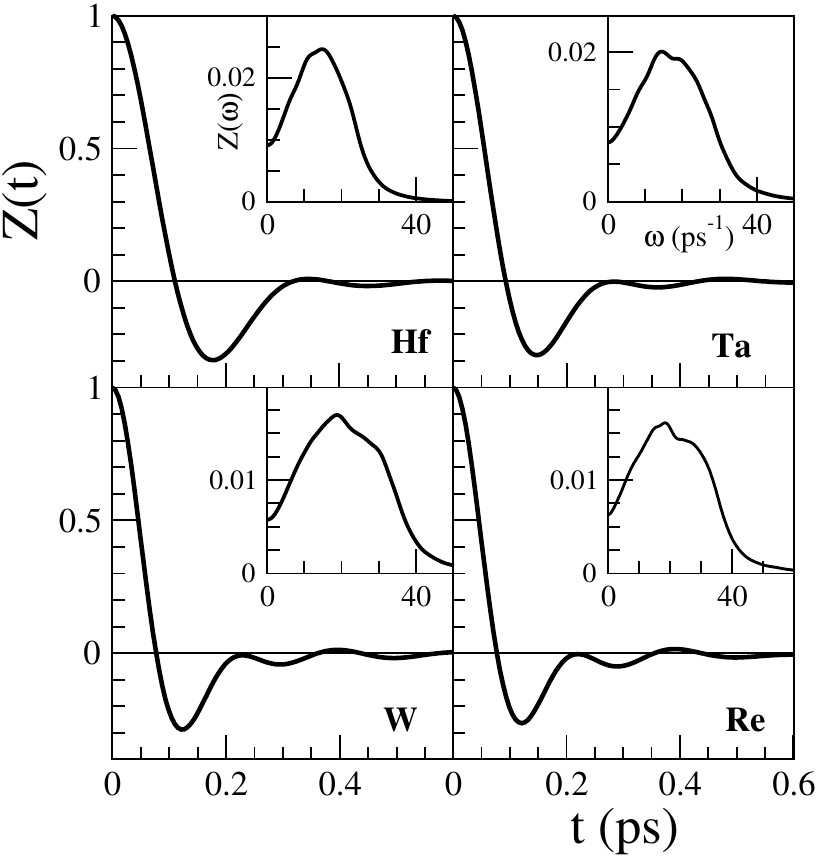}}
\caption{Normalized AIMD calculated velocity autocorrelation 
function of l-Hf, l-Ta, l-W and l-Re. 
The inset represents the corresponding  power spectrum   
$Z(\omega)$. }
\label{fcvfigA}
\end{figure}

\begin{figure}[!htb]
\centerline{\includegraphics[width=85mm,clip]{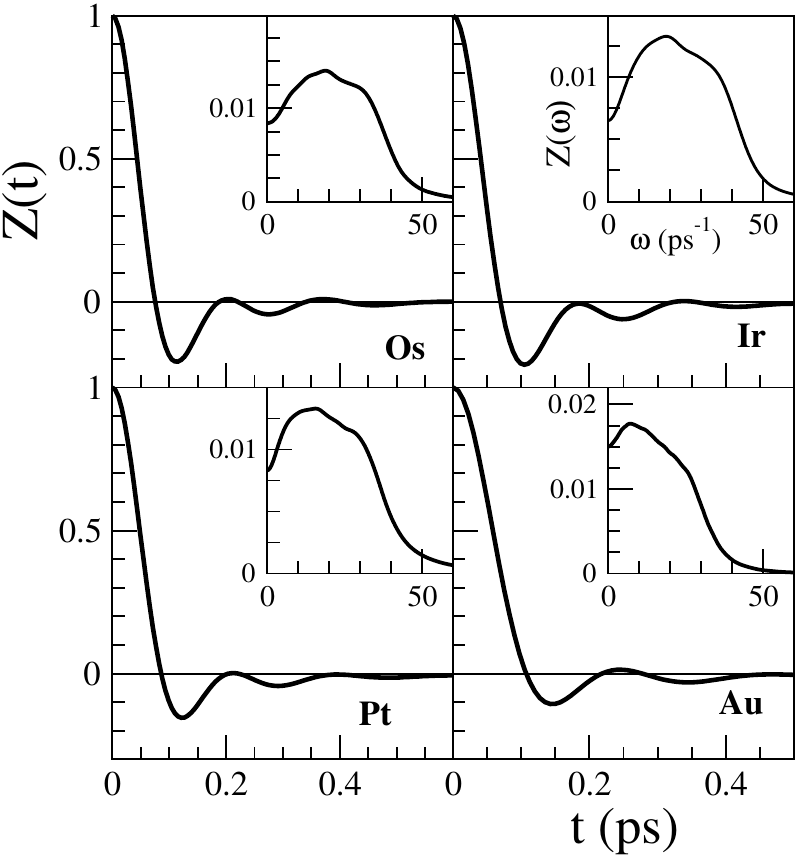}}
\caption{
	Same as the previous graph, but for 
	l-Os, l-Ir, l-Pt \cite{RioGG} and l-Au. }
\label{fcvfigB}
\end{figure}

 Figures \ref{fcvfigA}--\ref{fcvfigB} show the obtained 
AIMD results for $Z(t)$.  They display the  
typical decaying behavior with a 
marked first minimum related to the rebounding of the tagged atom
against the cage formed by its near neighbors, followed by 
 relatively weak oscillations that damp towards zero 
 at longer times.
Notice that the number density plays an important role
in the single-particle dynamics: a higher density
leads to a faster cage effect when moving from Hf to Ir,
and a subsequent decrease in the density when moving towards
Au produces some slowing down of the backscattering
minimum.

The Fourier Transform (FT) of the $Z(t)$ into the frequency domain 
gives the associated power spectra, $Z(\omega)$ and these are  
plotted as insets in figures \ref{fcvfigA}--\ref{fcvfigB}.   
Notice that the shapes of the $Z(\omega)$ 
may display just a peak (l-Hf) or  a
peak followed by a more/less marked shoulder located at a
higher frequency. 
Similar qualitative results were obtained in  
previous AIMD studies of several 3$d$ and 4$d$ liquid 
metals \cite{Bea_3d,Bea_3d2,Gonzalez_4d}. 
Some authors \cite{AuGuarini,Taras18}   
have recently suggested the existence of a relation between the shape of 
the $Z(\omega)$ and the appearance of a special type of collective 
excitations.
We will further comment about this point in the 
following section where we discuss our obtained 
results in connection with the transverse currents.

The self-diffusion coefficient, $D$, was calculated by both 
the time integral of $Z(t)$ and from the slope of the mean 
square displacement of a tagged ion in the fluid; both routes 
lead to practically the 
same results, which are given in table \ref{dynamic}.  
We are not aware of any experimental 
self-diffusion data for the 5$d$ metals considered in 
this work. Nevertheless, we  included 
in table~\ref{dynamic} some estimates obtained from 
semiempirical expressions based on the modified Stokes-Einstein type formulae   
which relate the self-diffusion coefficient to other 
thermophysical magnitudes such 
as  density and viscosity~\cite{IG15}.


\begin{table}[!htb]

	\caption{The calculated values of  
the self-diffusion coefficient ($D$), adiabatic sound velocity ($c_s$) and  
	shear viscosity ($\eta$)  for the 
liquid 5$d$ transition metals at 
the thermodynamic states given in table \ref{states}.
        The sound velocities included in the second lines for l-Ta, l-W, l-Pt
	and l-Au correspond to experimental values, and the second and
	third lines for the other elements are different semiempirical
	estimations.
	The viscosities reported in the second and third lines for each element 
	are the measured values in different experiments.
	\label{dynamic}}
\vspace{2ex}
\begin{center}
\begin{tabular}{ccccccc}
\hline
	$\;\;\;$  & $D$ (\AA$^2$/ps) &  $c_s$ (m/s) & $\eta$ (GPa ps) \\ 
\hline
	Hf &  \; 0.332 $\pm$ 0.020 & 2380  $\pm$ 150 &  4.90 $\pm$ 0.25 &  \\
	&  0.418 \cite{IG15}  & 2631, 2559   \cite{Blairs07}  &  5.2 \cite{Ishi03} & \\
	&  & 3371   \cite{IG15}  &  & \\
	\hline
	Ta &  \; 0.378 $\pm$ 0.025  & 3180 $\pm$ 150  & 6.00 $\pm$ 0.30 & \\
	&  0.365\cite{IG15}  & 3303  \cite{IG15,Blairs07} & 8.6 \cite{IG15} & \\
	&  & & 8.8 $\pm$ 0.9 \cite{Ishi13} & \\
	\hline
	W & \; \; 0.307 $\pm$ 0.015 & 3890 $\pm$ 150  & 8.60 $\pm$ 0.40 &  \\
	&  0.530\cite{IG15} & 3279  \cite{IG15,Blairs07}    & 7.0 \cite{IG15} &  \\
	&  & & 8.5 $\pm$ 0.9 \cite{Ishi13} &  \\
	\hline
	Re & \;  \; 0.333 $\pm$ 0.020 & 3640 $\pm$ 150  & 7.60 $\pm$ 0.35  &   \\
	&  0.447\cite{IG15} & 3569 \cite{IG15}   & 7.9 \cite{IG15}  & &  \\
	&  & 2943, 2665 \cite{Blairs07}   & 9.9 $\pm$ 1.0 \cite{Ishi13}  & &  \\
	\hline
	Os & \;  0.395 $\pm$ 0.025 &  2860 $\pm$ 200   &  5.90 $\pm$ 0.30  &   \\ 
	&  0.420\cite{IG15} & 3335 \cite{IG15}   & 4.2 \cite{IG15} & \\ 
	&  & 2777, 2487 \cite{Blairs07}   & 7.0 $\pm$ 0.7 \cite{Ishi13} & \\ 
	\hline
	Ir & \; 0.227 $\pm$ 0.015 & 3620  $\pm$ 200 &  8.80 $\pm$ 0.40  & \\
	&  0.400\cite{IG15}   &  3230 \cite{IG15}  & 6.0, 7.0 \cite{Ishi12} & \\
	&  &  2643, 2416 \cite{Blairs07}  &  & \\
	\hline
	Pt & \; 0.270 $\pm$ 0.015 & 3310  $\pm$ 200 &  4.90 $\pm$ 0.25  & \\
	&  0.305, 0.426\cite{IG15}   &  3053 \cite{IG15,Blairs07}  & 4.82, 6.74 \cite{IG15,Ishi12} & \\
	\hline
	Au & \; 0.282 $\pm$ 0.015 & 2380  $\pm$ 200 &  3.50 $\pm$ 0.20  & \\
	&  0.243\cite{IG15}   & 2568  \cite{IG15,Blairs07}  & 5.37 \cite{IG15} & \\
\hline
\end{tabular}
\end{center}
\end{table}


\subsubsection{Collective dynamics}

\noindent
The collective dynamics of density fluctuations in a liquid 
	can be described by the  
intermediate scattering function, $F(\vq, t)$, defined as 

\begin{equation}
F(\vq, t) = \frac{1}{N} \Bigl \langle  \sum_{i,j=1}^N  
 \exp \; [i \; \vq \cdot (\vR_i(t) - \vR_j(t=0))]   
 \;\Bigr  \rangle ,
\end{equation}

\noindent 
whose time FT gives its frequency spectrum, known as the dynamic structure 
factor, $S(\vq,\omega)$, which can be directly measured by either inelastic neutron 
scattering (INS) or inelastic X-ray (IXS) experiments.

\begin{figure}[!htb]
\centerline{\includegraphics[width=0.5\textwidth,clip]{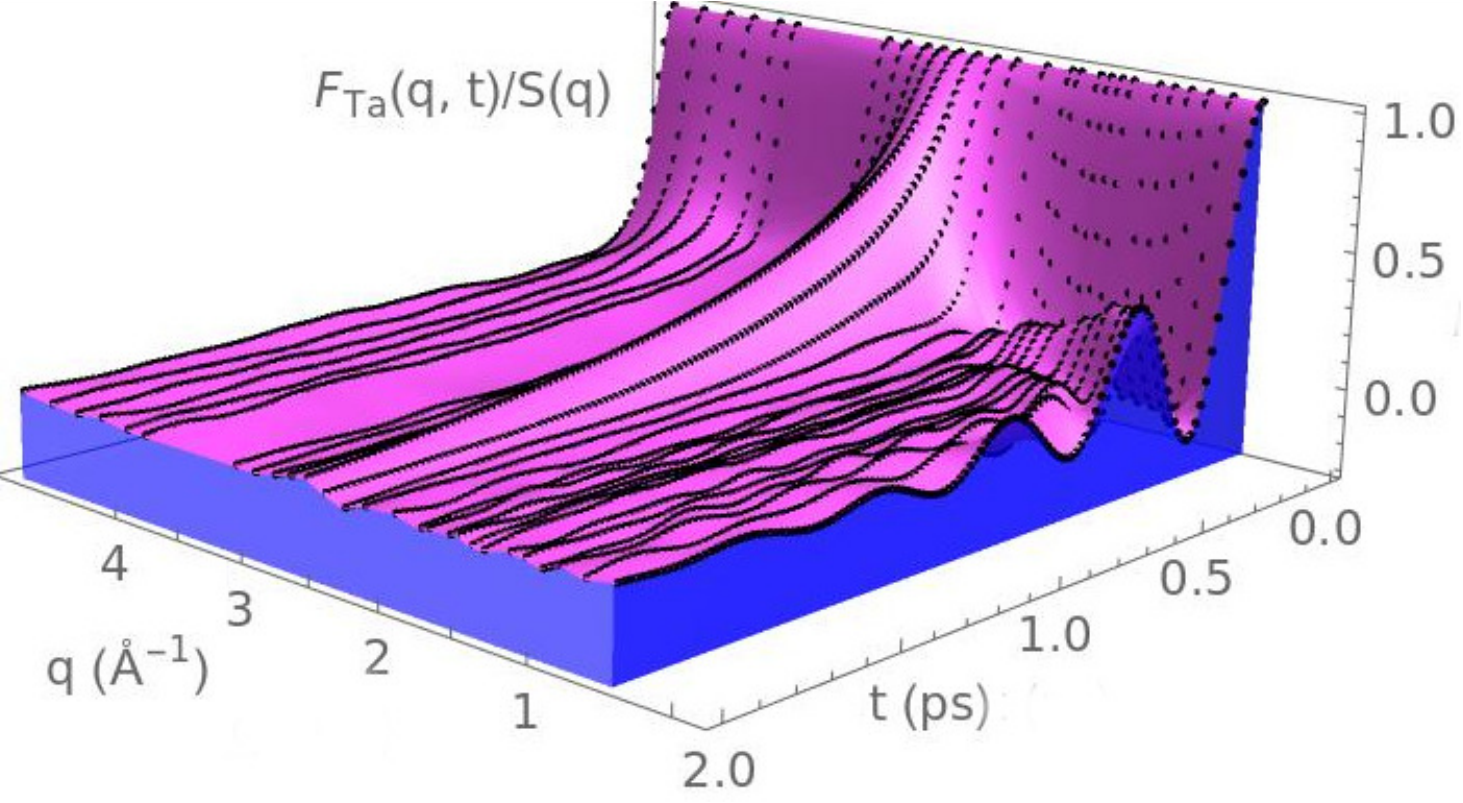}}
\caption{(Colour online) Intermediate scattering function, $F(q,t)/S(q)$, of 
l-Ta at $T=3350$ K  for 
several $q$ values. } 
\label{FktZr}
\end{figure}
\begin{figure}[!htb]
\centerline{\includegraphics[width=0.5\textwidth,clip]{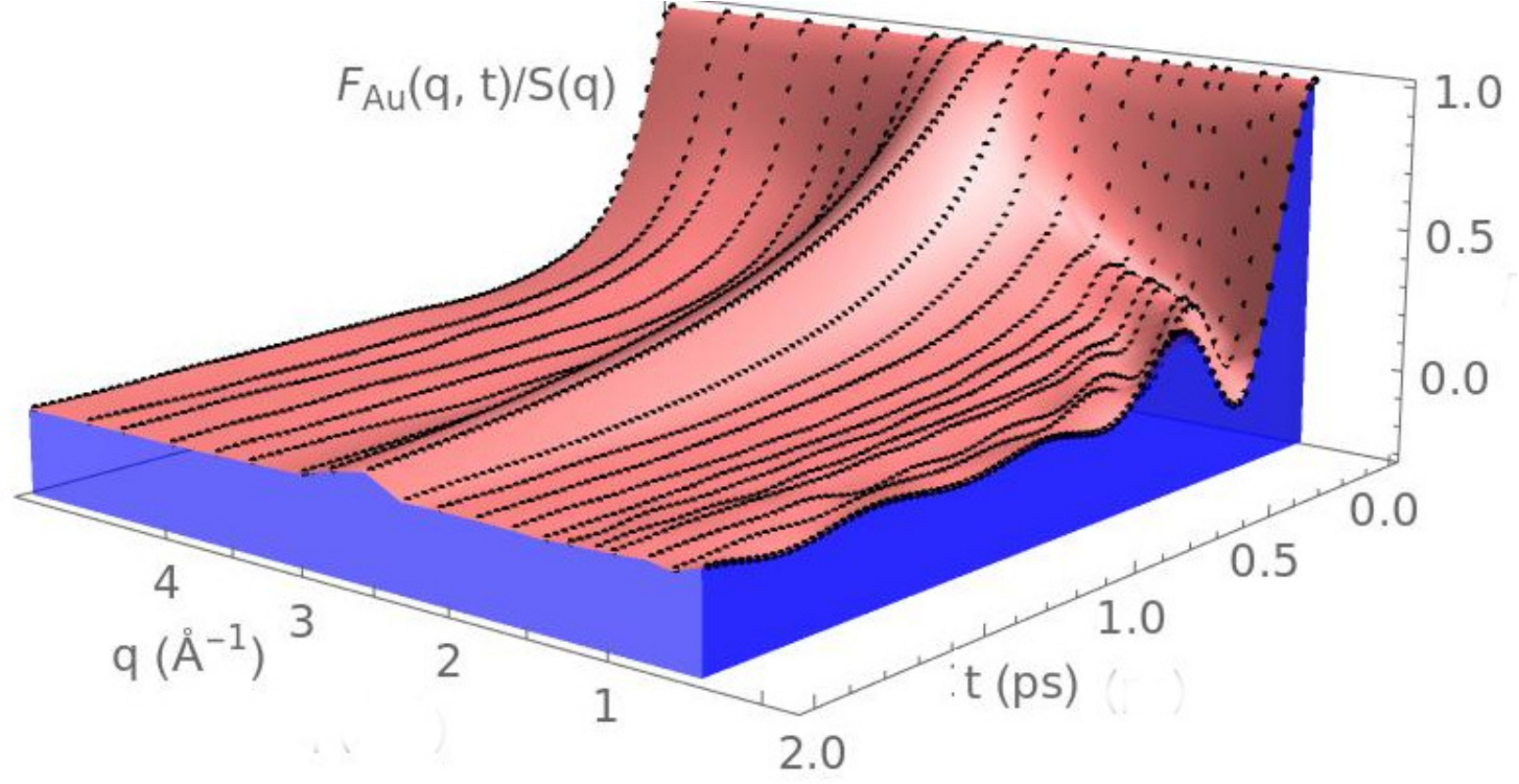}}
\caption{(Colour online) Intermediate scattering function, $F(q,t)/S(q)$, of 
l-Au at $T=1423$ K  for 
several $q$ values. } 
\label{FktRh}
\end{figure}

\begin{figure}[!htb]
\centerline{\includegraphics[width=0.5\textwidth,clip]{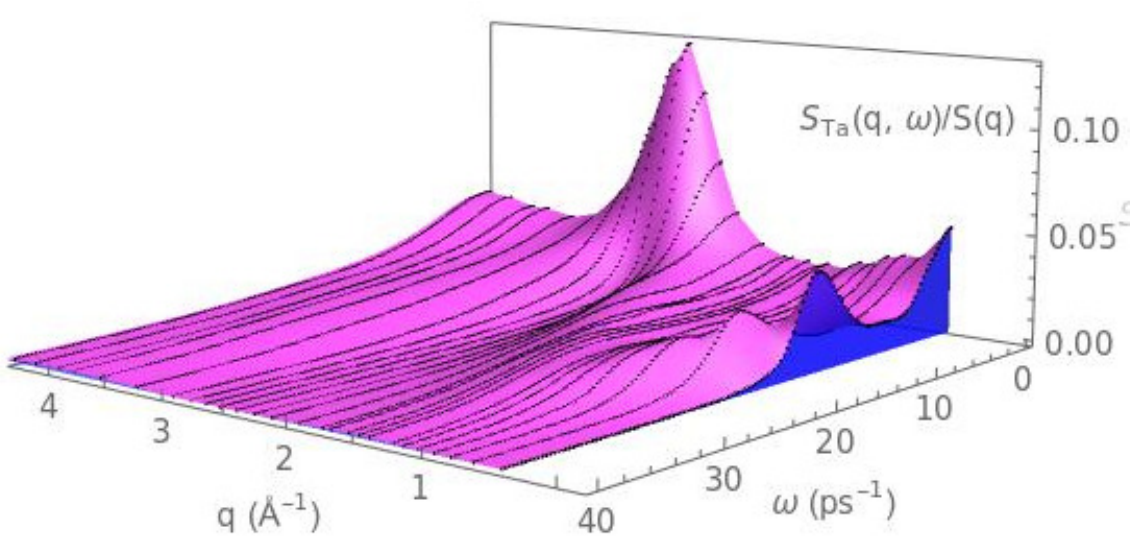}}
	\caption{(Colour online) Dynamic structure factors, $S(q, \omega)/S(q)$, of 
l-Ta at $T=3350$ K and  several $q$ values. } 
\label{sqwTa}
\end{figure}
\begin{figure}[!htb]
\centerline{\includegraphics[width=0.5\textwidth,clip]{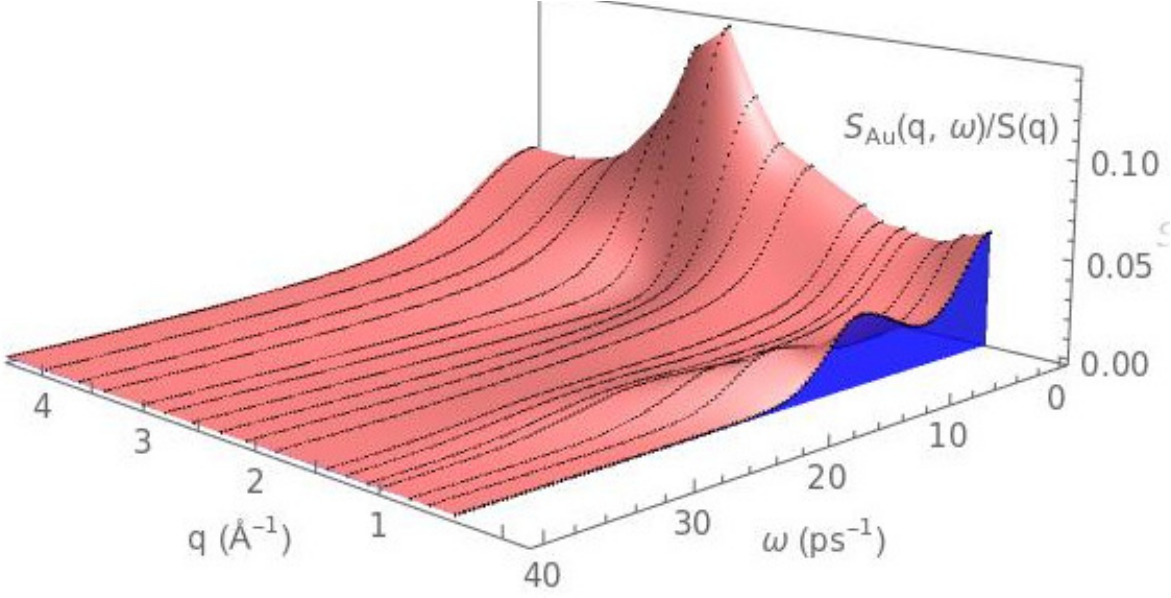}}
	\caption{(Colour online) Dynamic structure factors, $S(q, \omega)/S(q)$, of 
l-Au at $T=1423$ K and  several $q$ values. } 
\label{sqwAu}
\end{figure}

Figures \ref{FktZr}--\ref{FktRh} depict, for l-Ta and l-Au, 
the corresponding AIMD results for $F(q, t)$. The trends observed for these two metals are 
qualitatively similar for all the other metals: at small $q$'s, 
there is an oscillatory behavior that subsides with increasing $q$-values and  
 it has practically disappeared when $ q \approx (4/5) \, q_p$, where $q_p$ denotes 
	the position of the main peak of $S(q)$.
Then, for greater $q$ values ($q > q_p$), the $F(q, t)$ exhibit a monotonously 
decaying shape. 
The associated $S(q, \omega)$ are plotted  in 
figures \ref{sqwTa}--\ref{sqwAu}, and show,
up to $q \approx (3/5) \, q_p$, 
side-peaks which attest to 
the existence of collective density excitations; for greater $q$-values, the  
$S(q, \omega)$  display  a monotonously decreasing behavior. This same qualitative 
trend is exhibited by all the other $5d$ metals considered in this paper.

\begin{figure}[!htb]
\centerline{\includegraphics[width=0.5\textwidth,clip]{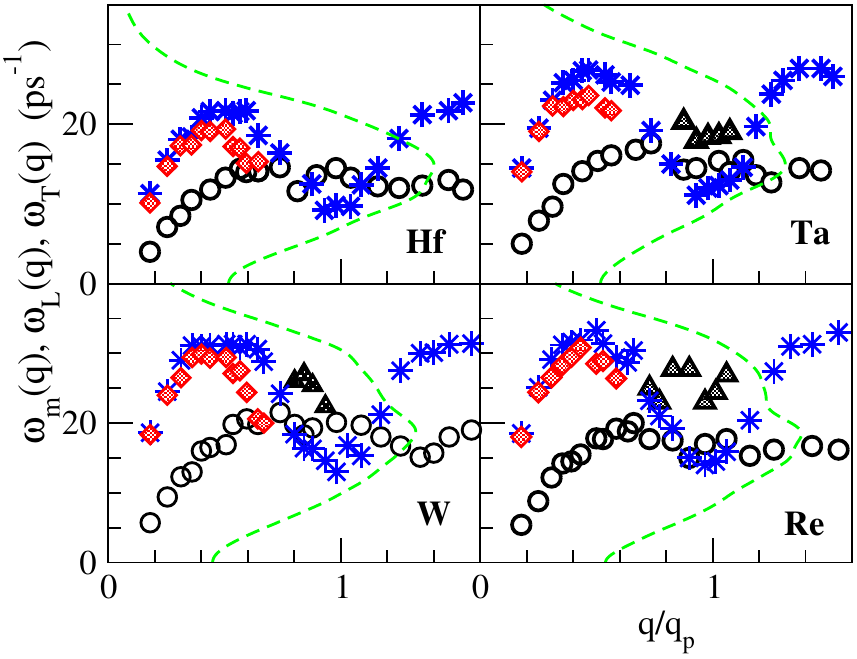}}
\caption{(Colour online) Dispersion relations for l-Hf, l-Ta, l-W and l-Re. 
Red diamonds and blue stars: longitudinal dispersion obtained from 
the AIMD results for the positions of the inelastic peaks
in the $S(q,\omega)$ and from the maxima in the spectra of the longitudinal
current, $C_L(q, \omega)$, respectively.
Open circles and triangles: transverse dispersion from the positions
of the peaks in the spectra $C_T(q, \omega)$. The green dashed line represents 
	the corresponding $Z(\omega)$. 
	}
\label{DisperLONTRANS-A}
\end{figure}

\begin{figure}[!htb]
\centerline{\includegraphics[width=0.5\textwidth,clip]{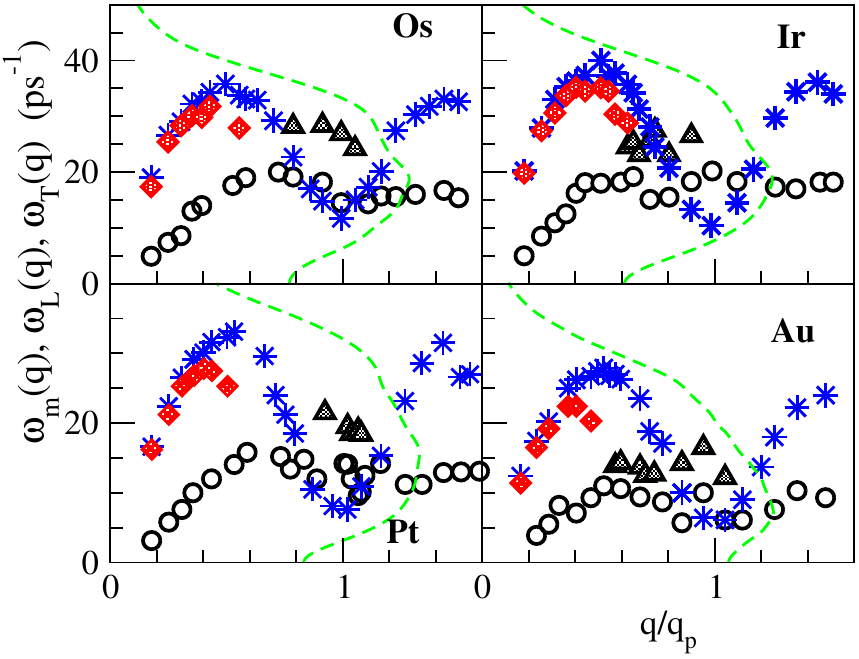}}
\caption{(Colour online) 
        Same as the previous graph, but for
l-Os, l-Ir, l-Pt and l-Au. }
\label{DisperLONTRANS-B}
\end{figure}

\begin{figure}[!htb]
\centerline{\includegraphics[width=0.6\textwidth,clip]{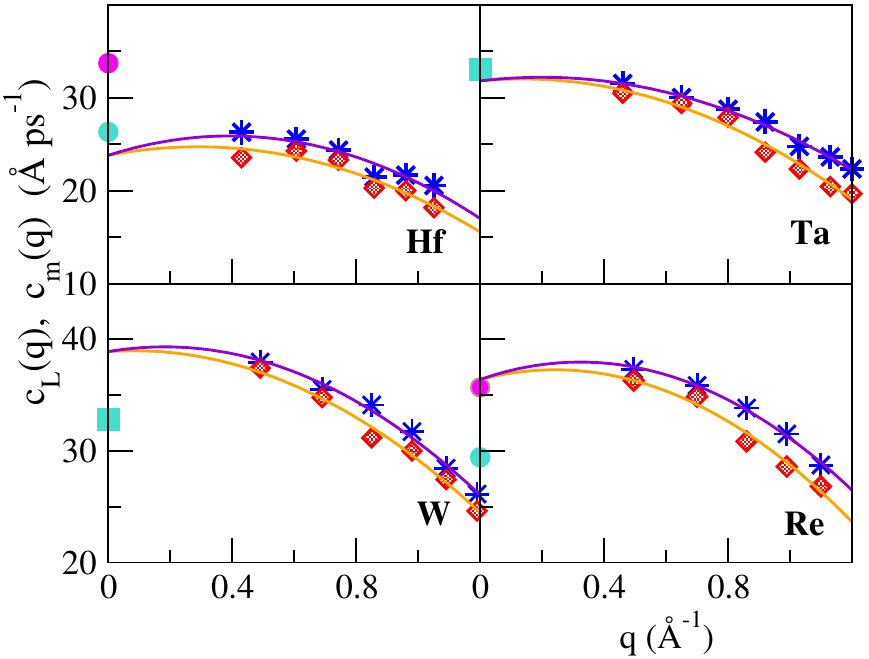}}
	\caption{(Colour online) Phase velocities as obtained from maxima in $S(q,\omega)$ (red diamonds) and
	from $C_L(q,\omega)$ (blue stars) for l-Hf, l-Ta, l-W and l-Re. 
	Lines are fits using a quadratic formula
	to both datasets with a common value at $q=0$. 
	Magenta circles denote
	estimations for the adiabatic sound velocities, $c_s$, from \cite{IG15}. Turquoise
	circles are estimations for $c_s$ from \cite{Blairs07} and turquoise squares
	are measured values of $c_s$ as reported in \cite{Blairs07}.
	}
\label{phvel-A}
\end{figure}

\begin{figure}[!htb]
\centerline{\includegraphics[width=0.6\textwidth,clip]{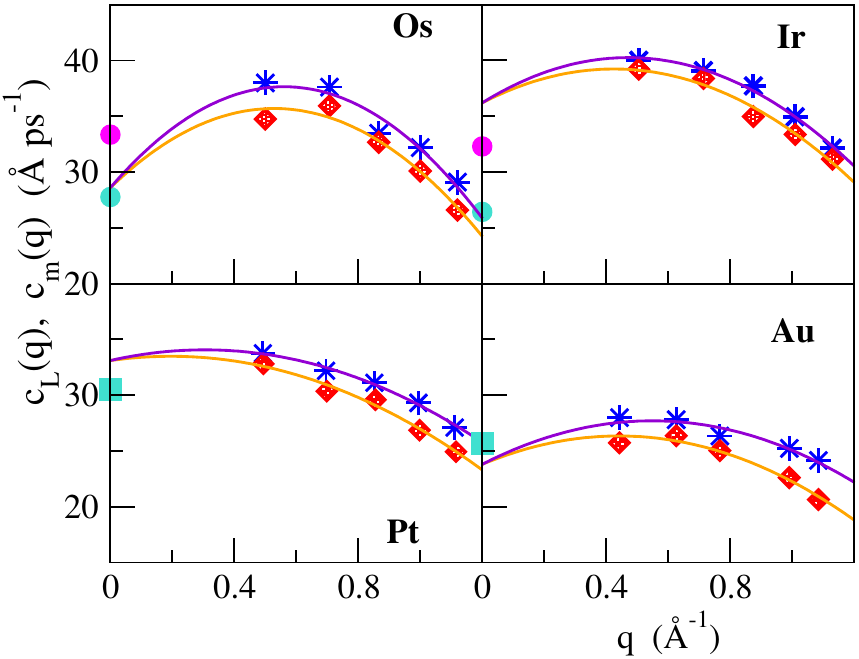}}
	\caption{(Colour online) Same as the previous figure, but for l-Os, l-Ir, l-Pt and l-Au.
	}
\label{phvel-B}
\end{figure}

For each metal, we  determined the frequency of the side-peaks as 
a function of the wavevector, namely the function $\omega_m(q)$ which 
represents the associated  dispersion relation of the density 
excitations and those are depicted in 
figures \ref{DisperLONTRANS-A}--\ref{DisperLONTRANS-B}. 
The corresponding phase velocities are defined as $c_m(q)=\omega_m(q)/q$.
These are plotted in figures \ref{phvel-A}--\ref{phvel-B} in the
region $q \leqslant 1.2$ \AA$^{-1}$.
In the long wavelength limit ($q \to$ 0), the phase velocity tends
towards the adiabatic speed of sound, $c_s$. The extrapolation of
the AIMD data to $q=0$ is subject to some uncertainty because
the minimum wavevector is restricted by the relatively small 
simulation box associated to the small number of particles 
affordable within the AIMD method.  Nevertheless, there are some
constraints that must be fulfilled in the long wavelength limit
that we have implemented, and are related to the phase velocities
associated to the longitudinal currents, which we define next.

The current due to the overall motion of the 
particles, 
\begin{equation}
\vj(\vq, t) =
\sum_{j=1}^{N} \vw_{j}(t) \; \exp [ \ri \vq \cdot
\vR_{j}(t) ],
\label{jiqt}
\end{equation}

\noindent 
is usually split into a longitudinal 
component $\vj_l(\vq, t) = (\vj (\vq, t)\cdot \vq) \vq /q^2$, 
and a transverse component $\vj_t(\vq, t)= \vj(\vq, t) - \vj_l(\vq, t)$.  
The longitudinal and transverse current correlation functions are obtained as

\begin{equation}
C_L(\vq, t)= \frac{1}{N} \langle
\vj_l(\vq, t) \cdot \vj^{ *}_l(\vq, t=0) \rangle \;
\label{CLjqt}
\end{equation}
and
\begin{equation}
C_T(\vq, t)= \frac{1}{2N} \langle
\vj_t(\vq, t) \cdot \vj^{ *}_t(\vq, t=0) \rangle.
\label{CTjqt}
\end{equation}

\noindent
The respective time Fourier Transforms (FT) give the 
associated spectra $C_L(\vq, \omega)$ and 
$C_T(\vq, \omega)$.

For any fixed $q$-value,  when the $C_L(q, \omega)$ is plotted as 
a function of $\omega$ we observe a maximum, and its associatted frequency, 
namely  $\omega_L(q)$, stands for 
the dispersion relation of the longitudinal modes.  
These are plotted in figures \ref{DisperLONTRANS-A}--\ref{DisperLONTRANS-B} where 
it is observed that their shape shows a  
trend  similar to that already observed 
in other liquid metals \cite{ScopigRev}.    
The associated phase velocities, $c_L(q)=\omega_L(q)/q$, are also plotted in figures
\ref{phvel-A}--\ref{phvel-B}, and must tend towards the same value, $c_s$, as the
$c_m(q)$ when $q\to 0$. We  performed a quadratic fit $c(q)=c_s+aq+bq^2$
to the phase velocities $c_m(q)$ and $c_L(q)$ in the $q$ range shown in
figures \ref{phvel-A}--\ref{phvel-B} imposing a common value $c_s$ at $q=0$.
In all cases we obtain a positive value of the coefficient $a$, an effect 
named as positive dispersion that is quite usual in liquid metals near their
melting points at ambient pressure, and was rationalized by Bryk and 
coworkers resorting to theoretical models based on the generalized collective
modes approach \cite{Bryk-LJ}. 
The results obtained for $c_s$ are given in  
table \ref{dynamic}, which  also includes for comparison
the available experimental (l-Ta, l-W, l-Pt, l-Au)~\cite{Blairs07} and 
semiempirical (l-Hf, l-Re, l-Os, l-Ir) \cite{IG15,Blairs07} data. 
These latter data are the estimates obtained from some semiempirical 
formulae \cite{Blairs07} which relate the adiabatic sound velocity 
to other thermophysical magnitudes 
such as surface tension, density and melting temperature. 
Overall, we observe a good agreement between the AIMD results and 
the experimental/semiempirical data, 
with the greatest discrepancies observed for l-W  and l-Ir.

The $C_T(q,t)$ has a mathematical structure which evolves from a 
gaussian function, in both $q$ and $t$, at the free particle ($q \to \infty$) limit,  
towards a gaussian in $q$ and  exponential in $t$ at the hydrodynamic 
limit ($q \to 0$). Although in both limits, the $C_T(q,t)$ is always positive, 
for intermediate $q$-values it generally displays noticeable oscillations 
around zero \cite{Balubook,GGLS,GGLS2,HanMc}, related to the propagation of
damped shear waves.

Its associated spectrum $C_T(q, \omega)$, when plotted as 
a function of $\omega$,  shows within some $q$-range, clear peaks 
which indicate the frequencies of the propagating shear waves. 
In the present AIMD simulations, the lowest attainable $q$ value, $q_{\rm min}$, 
is already outside the hydrodynamic region and the associatted 
 $C_T(q=q_{\rm min}, \omega)$ displays a peak which attests to 
the propagation of shear waves. 
With increasing $q$-values,  we obtain that  
the  frequency of the peak, $\omega_T(q)$, increases, reaches a 
maximum  at $q \approx (2/3) \, q_p$, and after a slight initial decrease, the 
$\omega_T(q)$ remains practically constant  until 
$q \approx 3.0 \, q_p$ when  the peaks disappear. 
Furthermore, we also find, for some 
metals, another peak with a higher frequency, although its appearance lasts 
over a much smaller $q$-range located around the position of the main peak of $S(q)$. 
These features are depicted in 
figures \ref{DisperLONTRANS-A}--\ref{DisperLONTRANS-B} where we  plotted 
the high and low-frequency dispersion relations for the transverse modes. 
Notice that the two branches occur in all of the systems except l-Hf.

A few years ago, Bryk  et al. \cite{BRSS15} reported for the first time 
the presence of such a second, high-frequency, transverse branch in an AIMD study of l-Li at  
high pressures,  and this result was later confirmed by similar findings in AIMD calculations 
of high pressure  l-Na and l-Fe \cite{MGGNa-Fe,MGGNa-Fe2}. Although it was initially considered to be a 
feature induced by the high pressure state,  subsequent studies revealed 
a second high-frequency transverse branch  in  
AIMD studies of  l-Tl, l-Pb, l-Zn, l-Sn, and some liquid 3$d$ and 4$d$ transition metals 
at ambient pressure \cite{TDJ19,TDJ19b,Taras18,Bea_3d,Bea_3d2,Gonzalez_4d,BeaLuis1718,BeaLuis1718b}. 
A possible explanation for the appearance of this second transverse branch 
was put forward in terms of   
mode-coupling (MC) ideas \cite{BeaLuis1718,BeaLuis1718b}.

Recently, Bryk and coworkers \cite{TDJ19,TDJ19b,Taras18}   pointed out to the 
simultaneous appearance of both the  
high frequency transverse modes  and the 
presence of another,  high frequency peak/shoulder, in the
spectra of the VACF, namely $Z(\omega)$. 
They observed this feature in AIMD studies 
of l-Tl and l-Pb; moreover, they found a definite correlation
between the frequency peaks of the $Z(\omega)$  and the average 
frequencies,  in the $q$ region between $q_p/2$ and $q_p$, 
of both high- and low frequency transverse branches. 
The present AIMD simulation results ratify this connection 
between the structure of the $Z(\omega)$ and the existence of 
one/two  transverse dispersion branches. 
According to  
figures \ref{DisperLONTRANS-A}--\ref{DisperLONTRANS-B}, 
we find that l-Hf has a $Z(\omega)$ with one peak  and 
the associated transverse dispersion relation displays 
just one, low-frequency, branch.  
However, for the other 5$d$ metals we observe that the 
associated $Z(\omega)$  displays one peak along with 
a (higher-frequency) shoulder/peak  
and the corresponding transverse dispersion shows two branches. 
We note that a similar correspondence was found in AIMD simulation studies 
of several liquid 3$d$ and 4$d$ transition metals~\cite{Bea_3d,Bea_3d2,Gonzalez_4d}.
In principle, such a connection should not be unexpected because 
several years ago Gaskell and Miller \cite{GasMi78,GasMi78-1,GasMi78-2}  used the MC theory to 
formulate a theory of the VACF  based on  
contributions arising from the coupling of the single particle motion to the collective
longitudinal and transverse currents. However, a more detailed account 
about the way these correlations are established is still lacking.

From the AIMD results for the $C_T(q, t)$, we  also  
evaluated \cite{GGLS,GGLS2,PalmerBaBroJedVa,PalmerBaBroJedVa2} the associated shear viscosity 
coefficient, $\eta$, and in  table \ref{dynamic} we report 
our results along with the available experimental data. 
Fortunately, despite the technical challenges caused by 
these 5$d$ refractory metals,  their respective 
viscosities were measured by 
means of levitation techniques \cite{IG15,Ishi03,Ishi13,Ishi12} with an 
uncertainty of around 10\%. We obtain a fair agreement  with experimental data 
even though we observe, for some metals, noticeable differences among those 
data.

\begin{figure}[!htb]
\centerline{\includegraphics[width=0.45\textwidth,clip]{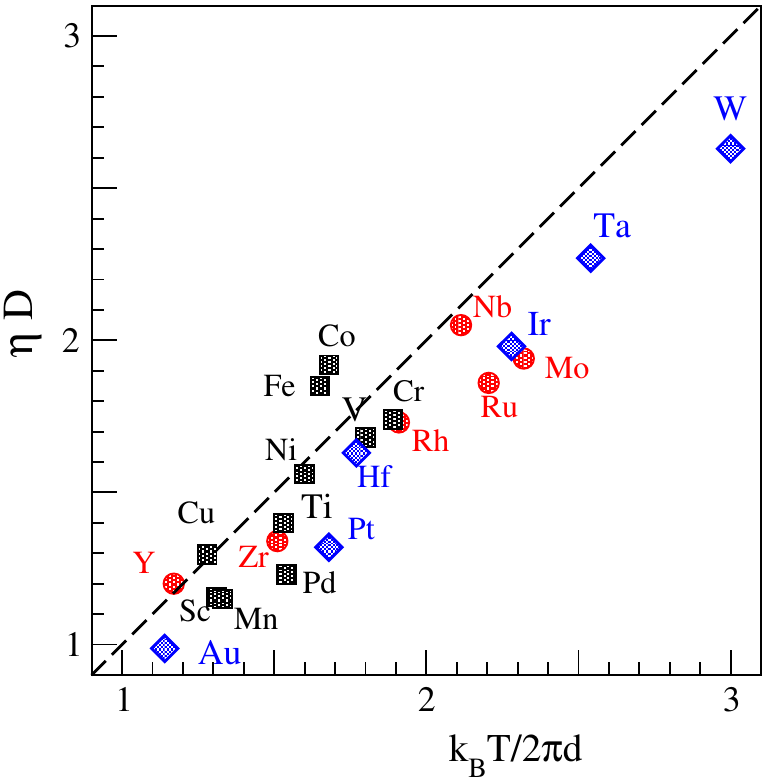}}
	\caption{(Colour online) Stokes-Einstein relation. Black squares: 3$d$ metals.
        Red circles: 4$d$ metals. Blue diamonds: 5$d$ metals.
	}
\label{Stokesfig}
\end{figure}

The study of transport properties in metallic liquids has often
resorted to the Stokes-Einstein (SE) relation which 
establishes a connection 
between the self-diffusion coefficient $D$ of a particle  with a diameter 
$d$ moving in a liquid of viscosity $\eta$.
Although the SE relation was originally aimed at 
brownian particles and, therefore, was 
derived by using purely macroscopic
considerations, it was found to work rather well for many 
monoatomic liquids \cite{HanMc}.  
In the slip condition, the SE relation reads as 
$\eta D=k_{\text{B}} T/2\piup d$ and it has often been used 
to estimate $\eta$ (or $D$) by identifying
$d$ with the position of the main peak of $g(r)$.
In the specific case of MD simulations, as the evaluation of the 
self-diffusion coefficient takes much less computation time than 
the calculation of the shear viscosity, then it has become a common 
practice to use the SE relation to get an estimate of the 
shear viscosity.  
Moreover, the SE relation has also been used in the derivation of 
several semiempirical formulae connecting, in terms of other 
thermophysical magnitudes,   
the shear viscosity  and the self-diffusion coefficient. 

By using the present AIMD results for the diffusion coefficients and the
viscosities of the liquid $5d$ metals we have analyzed the accuracy of the
SE relation. The results are plotted in figure~\ref{Stokesfig}, along 
with similar AIMD-based data for the liquid $3d$ and $4d$ metals
\cite{Bea_3d,Bea_3d2,Gonzalez_4d}.

\begin{figure}[!htb]
\centerline{\includegraphics[width=0.4\textwidth,clip]{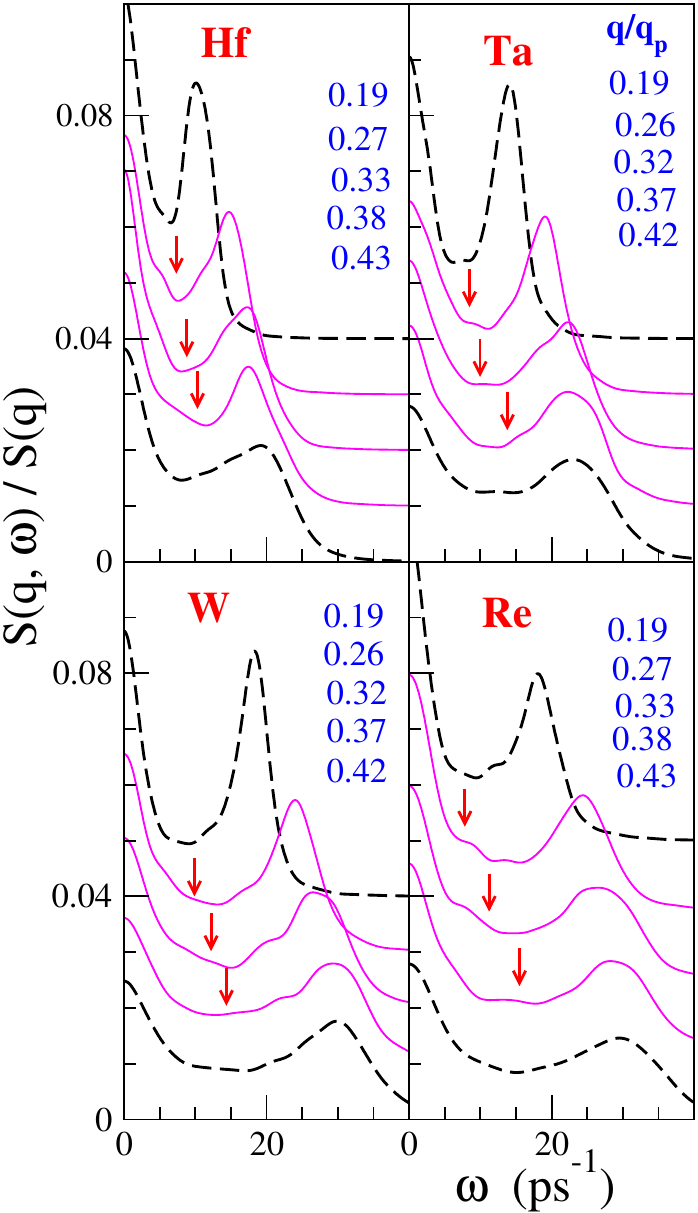}}
\caption{(Colour online) AIMD calculated dynamic structure factors $S(q,\omega)/S(q)$ 
for l-Hf, l-Ta, l-W and l-Re at  various (top to bottom)  $q/q_p$ values. 
The vertical scales are offset for clarity. The 
arrows point to the locations of the   
	maxima in the $C_T(q, \omega)$.
}
\label{SqwTRANS-A}
\end{figure}
\begin{figure}[!htb]
\centerline{\includegraphics[width=0.4\textwidth,clip]{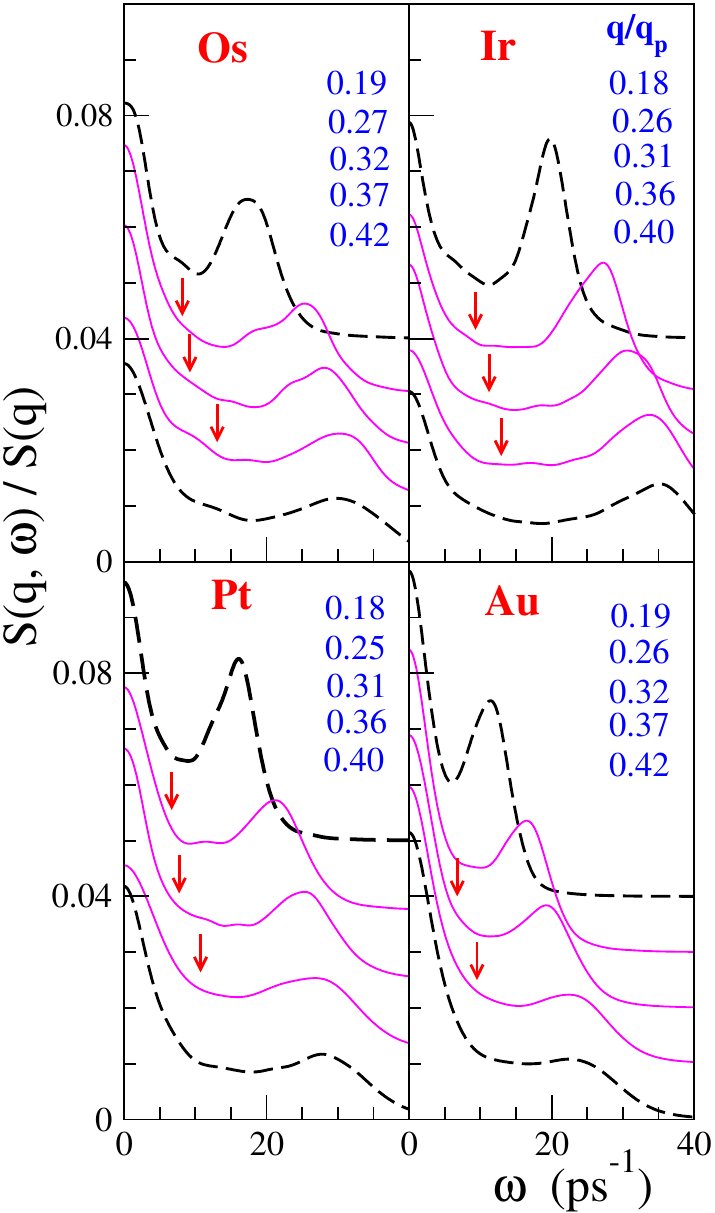}}
\caption{(Colour online) Same as the previous graph, but for 
for l-Os, l-Ir, l-Pt and l-Au. 
}
\label{SqwTRANS-B}
\end{figure}

In the last few years, another topic of interest 
has been the appearance of 
small, weak shoulders
in the dynamic structure factor of liquid metals at 
wavevectors in the wavevector region around $q_p/2$ and 
the frequency region between the quasielastic and 
inleastic peaks. 
These features were first noticed in an IXS measurement of the $S(q, \omega)$ 
of l-Ga near melting \cite{HosoTGa} and were 
corroborated by an OF-AIMD simulation study. They were interpreted as 
transverse-like low-energy excitations and were also  found 
in both measurements and AIMD simulation 
studies  \cite{Bea_3d,Bea_3d2,Gonzalez_4d,RGGNi,RGGNi2,HosoTSn,HosoTCuFe} for a range of liquid metals. 

Figures \ref{SqwTRANS-A}--\ref{SqwTRANS-B} show 
our AIMD simulation results 
for $S(q, \omega)$ at several $q$-values  
below $q_p/2$. We established that for most of 
these liquid metals  (except l-Pt and l-Au), 
the corresponding $S(q, \omega)$ shows 
the appearance of some weak shoulders whose  energies are close to those 
corresponding to the (low-frequency) peaks in the transverse current spectra. 
The physical origin of these low energy excitations in 
liquid metals is still a moot point because no sound theoretical scheme 
has provided an explanation as to how transverse currents could couple
to longitudinal excitations. Other possible interpretations, such as 
relaxing (non-propagating) terms or non-lorentzian (asymmetry) terms 
\cite{Montfrooij,Baron},
or as kinetic heat waves~\cite{Bryk-heat} have also been proposed.

\subsection{Electronic properties: density of states}

\begin{figure}[!htb]
\centerline{\includegraphics[width=0.7\textwidth,clip]{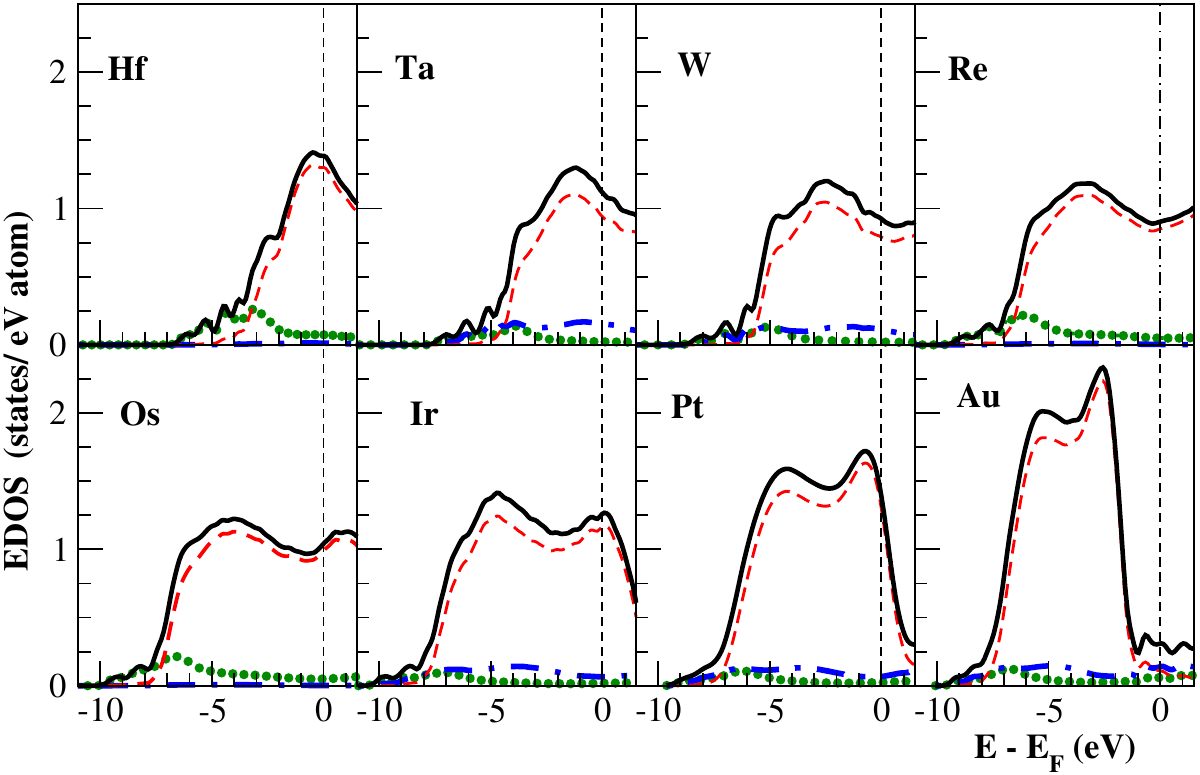}}
\caption{(Colour online) Total electronic density of states (black line) for several 
	$5d$ transition metals. 
The angular momentum decomposition of the DOS in 
$s$ (green dotted line), 
$p$ (blue dashed-dotted line) and  $d$ (red dashed line).
	Data for l-Pt are taken from \cite{RioGG}.
}
\label{dosfig}
\end{figure}

We  also calculated the 
partial and total electronic density of
states, $n(E)$. This was obtained from the
self-consistently determined eigenvalues, 
averaged over four ionic configurations
well separated in time ($\approx$ 15.0 ps), sampling 
the Brillouin zone with 
a mesh of $2\times2\times2$ points. 

Figure \ref{dosfig} shows the obtained results for the
electronic partial and total $n(E)$ associated to the 
outer valence electrons.  
The $n(E)$ is dominated by the 5$d$ states which are progressively filled 
with increasing atomic number. This is reflected in the  
increase of the height and/or width of the $n(E)$ as the occupation rises.
Additional (and smaller) contributions 
from the $6p$ and $6s$ states are also observed. 

The integrated partial densities of states up to the Fermi level indicate
the ocupations of the corresponding $s$, $p$ and $d$ bands. Such ocupations
are not integer in general, and add up to the total number of valence 
electrons per atom. Neglecting the lower lying $5s$ and $5p$ atomic states, the
occupations of the bands are reported in table \ref{dosoccup}. These data
may be of value for some models that treat within different
approximate schemes the $s$, $p$ electrons and the $d$ electrons in order
to obtain effective interatomic pair potentials \cite{Bhuiyan-22,Bhuiyan-22b,Bhuiyan-22c}.

\begin{table}[!htb]
\caption{
Occupations of the $s$, $p$ and $d$  bands in the liquid $5d$ metals 
considered.
\label{dosoccup}} 
\vspace{2ex}
\begin{center}
\begin{tabular}{ccccccccc}
	& Hf & Ta & W & Re & Os & Ir & Pt & Au  \\
\hline
	$6s$ & 0.8 & 0.3 & 0.3 & 1.0 & 1.0 & 0.4 & 0.4 & 0.5 \\
	$6p$ & 0.0 & 0.8 & 0.9 & 0.0 & 0.0 & 1.0 & 0.9 & 1.0 \\
	$5d$ & 3.2 & 3.9 & 4.8 & 6.0 & 7.0 & 7.6 & 8.7 & 9.5 \\
	$sp$ & 0.8 & 1.1 & 1.2 & 1.0 & 1.0 & 1.4 & 1.3 & 1.5 \\
\hline
\end{tabular}
\end{center}
\end{table}

\section{Conclusions}
\label{conclusions}

A range of static, dynamic and electronic properties of several liquid  
5$d$ transition metals have been calculated by using an 
{\it ab initio} simulation  method. 
Although there were some previous studies for l-Ta and l-Au, 
this is the first comprehensive AIMD study performed on their 
liquid state properties near melting, except l-Pt, whose results
are also included here for completion. 

As for the static structure, we obtain a  
good agreement with the available experimental data  in those two 
metals (l-Pt and l-Au) for which the comparison can be performed.
The calculated $S(q)$ displays an asymmetric
shape in its second peak with a marked 
shoulder in l-Hf, l-Ta, l-W and l-Re. This feature  
is connected with the existence 
icosahedral short-range order in the liquid. 
A subsequent analysis based on the CNA method shows that these four metals 
have the greatest abundance of five-fold type structures.

The AIMD dynamic structure 
factors, $S(q,\omega)$, show side-peaks which are 
indicative of collective density excitations. 
Furthermore, we have also found that the $S(q, \omega)$ associated to 
most of these metals (except l-Pt and l-Au) exhibit, within some 
$q$-range, some kind of excitations which have the  features similar to 
the transverse-like excitation modes  
found in IXS and INS experimental data for 
several other liquid metals.

The AIMD transverse current correlation functions, $C_T(q, t)$, display 
oscillations around zero, while their spectra, $C_T(q, \omega)$, show peaks which 
point to the existence of shear waves. Except l-Hf, we have found that the  
respective transverse dispersion relations exhibit two branches.
These  results support  the proposed connection between
the structure of the spectra of the VACF and the existence of one/two
transverse dispersion branches.

From the previous time correlation functions, we have also evaluated some 
transport coefficients, namely the  
self-diffusion, adiabatic sound velocity and shear viscosity coefficients. 
Taking into account the scarcity of data concerning most of these coefficients 
we expect that the present results will be helpful.  Finally, the previous 
results for self-diffusion and shear viscosity have been used to analyze the 
accuracy of the Stokes-Einstein relationship for these liquid metals.

\section*{Acknowledgements}

We acknowledge the support of the Spanish Ministry 
of Economy and Competitiveness (Project PGC2018-093745-B-I00), partly 
supported by european FEDER funds.

\newpage
\ukrainianpart

\title{Першопринципні дослідження статичних, динамічних та електронних властивостей деяких рідких
	5d перехідних металів поблизу точок плавлення}
\author{Д. Х. Гонзалес,   Л. Е. Гонзалес}

\address{Факультет теоретичної фізики, Університет Вальядоліду, 47011 Вальядолід, Іспанія}

\makeukrtitle

\begin{abstract}
	Представлено результати дослідження статичних та динамічних властивостей низки
	рідких 5$d$ перехідних металів в термодинамічних умовах в околі відповідних точок плавлення. 
	Дослідження проведено шляхом першопринципного моделювання методом молекулярної динаміки
	в рамках теорії функціоналу густини. 
	Представлено результати для статичних структурних факторів і парних функцій розподілу; 
	також аналізується локальний ближній порядок в рідких металах. 
	Обчислено як одночастинкові, так і колективні динамічні характеристики. 
	Отримана динамічна структура свідчить про флуктуації густини. Також отримано відповідне дисперсійне співвідношення і низку результатів для поздовжних і поперечних  спектральних функцій струмів разом з пов'язаною
	з ними дисперсією колективних збуджень. Для деяких металів виявлено існування двох гілок  
	поперечних колективних збуджень в області поблизу головного піку структурного фактора. 
	Проведено розрахунок деяких коефіціентів переносу.
	\keywords рідкі метали, перехідні метали, першопринципні розрахунки
\end{abstract}

\label{last@page}

\end{document}